\documentclass[aps,twocolumn,preprintnumbers,groupedaddress,nofootinbib,pra]{revtex4-2}

\usepackage{commath,amssymb,mathtools}
\usepackage{hyperref}
\hypersetup{
  colorlinks   = true, 
  urlcolor     = blue, 
  linkcolor    = black, 
  citecolor    = red 
}

\begin{document}

\title{A remarkably simple covariant graviton propagator in Anti-de Sitter spacetime}

\author{Radu N. Moga, Kostas Skenderis}
\affiliation{STAG Research Centre \& Mathematical Sciences, University of Southampton, Highfield, SO17 1BJ, Southampton, UK}

\begin{abstract}
    We present remarkably simple covariant expressions for the graviton and ghost propagators in Anti-de Sitter (AdS) spacetime valid in any spacetime dimension. In gravity there is a $2$-parameter family of covariant gauge-fixing conditions and the simplification occurs for a special choice of these parameters. In this gauge the graviton propagator satisfies $\nabla^\mu\mu\nabla^\nu\mu \, G_{\mu\nu,\alpha'\beta'}(\mu)=0$, where $\mu$ is the geodesic distance between two points, and this condition implies an improved infrared behavior. This gauge choice is not possible in flat spacetime. 
\end{abstract}

\maketitle

In this Letter we present a particularly simple graviton propagator for Anti-de Sitter (AdS) spacetime. While a fully satisfactory quantum framework for gravity with such asymptotics is still work in progress, at low energies the theory is well-described by General Relativity with a negative cosmological constant. Within this approximation, quantum effects may be computed using perturbative quantum gravity. The main ingredient for such computations is the graviton propagator, defined as the inverse of the differential operator appearing in the kinetic term of the metric fluctuation in the Einstein-Hilbert (EH) action.

Graviton fluctuations around flat spacetime have been studied long ago, see \cite{Feynman:1963ax,DeWitt:1967yk,DeWitt:1967ub,DeWitt:1967uc,tHooft:1974toh} for a selection of seminal works. Our focus here is on fluctuations around AdS. One of the main motivations is the AdS/CFT correspondence \cite{Maldacena:1997re,Gubser:1998bc,Witten:1998qj}, which provides a window into a non-perturbative definition of quantum gravity. The correspondence is currently best understood at tree-level, which corresponds to the leading large $N$ behavior in the 't Hooft large $N$ limit of the dual CFT. However, to further explore the insights encoded in this duality it is necessary to go beyond the classical approximation and study quantum fluctuations. This would allow one to study subleading corrections in the large $N$ expansion. Since the metric is dual to the energy-momentum tensor such bulk quantum corrections would capture universal $1/N^2$ corrections. The case of positive cosmological constant is also interesting. The maximally symmetric solution, the de Sitter (dS) spacetime, provides an excellent approximation for the period during which our universe underwent primordial inflation and may also be of relevance for the current dark energy era. Thus, beyond the importance on theoretical grounds, quantum gravitational effects in dS may also be of phenomenological interest. We will comment on this case in the concluding paragraph.

Quantum computations on curved spacetimes are significantly more difficult than those in flat spacetime, one of the main reasons being that propagators in curved spacetimes (when they can actually be computed explicitly)  are given by complicated expressions often involving special functions. This makes the computation of Feynman diagrams very difficult, as the integrals involved are intractable. In this Letter, we address this problem in AdS by presenting a particularly simple covariant expression for the graviton propagator. 

Because gravity is diffeomorphism invariant, which is a gauge symmetry, in order to define the graviton propagator one needs to gauge fix by following the Faddeev–Popov procedure and introducing corresponding ghost fields into the action. Equivalently, one can proceed by following the prescription of Becchi-Rouet-Stora-Tyutin (BRST) quantization and start directly from the BRST invariant action.

The AdS graviton propagator has been computed previously in non-covariant gauges in \cite{Liu:1998ty,Raju:2011mp}, whereas only its gauge independent parts have been computed in a covariant gauge \cite{DHoker:1999bve} (see, however,  \cite{Kolar:2023mkw}). While these parts were enough for the tree-level computations that were performed at the time, one needs the full expression for the graviton propagator, including its gauge dependent parts, in order to carry out loop-level computations. There is also relevant literature discussing the dS graviton propagator. The propagator in dS has been obtained in various gauges, including covariant \cite{Allen:1986ta,Allen:1986tt,Antoniadis:1986sb,Higuchi:2001uv,Miao:2011fc,Kahya:2011sy,Mora:2012zi,Frob:2016hkx} and non-covariant \cite{Tsamis:1992zt,Tsamis:1992xa,Higuchi:2002sc,Glavan:2019msf,Glavan:2025azq} ones (see also references therein). The de Sitter case is more subtle and there is an ongoing debate about dS covariance, infrared (IR) divergences, and how to deal with them, see, for example \cite{Higuchi:2011vw, Morrison:2013rqa, Miao:2013isa}. None of these issues, however, are relevant to the AdS case.

There are various advantages and disadvantages associated with different gauge choices and corresponding representations for the propagator, see \cite{Moga:2025gdy} for a recent discussion regarding the spin-$1$ case. Here we will focus on obtaining a covariant expression. One of the reasons for this is our interest in generalizing the recent results of \cite{Banados:2022nhj} to the gravitational setting. In \cite{Banados:2022nhj} position space techniques were employed for scalars and AdS isometries were used in an essential way to simplify the computation of loop-level integrals arising from Witten diagrams. Thus, a covariant expression for the AdS propagator would be particularly useful. 

In most of the existing literature, the expression for the graviton propagator is broken down into its various parts, {\it i.e.}\ transverse traceless, vector, and scalar parts. While this is {\it a priori} sensible, experience with the photon propagator \cite{Moga:2025gdy} suggests that linear combinations of the various parts may generate unexpected cancellations leading to simple expressions for special gauges. With that in mind, unlike previous literature, we will not disentangle various parts of the propagator, but rather follow the approach presented in \cite{Moga:2025gdy} for the photon propagator.

The rest of this Letter is organised as follows. We will first set-up our computation for the graviton propagator in the most general covariant gauge. We will present the solution for the propagator in a special gauge for which the expressions simplify dramatically and comment on the reasons this gauge is special.  We will then finish with a discussion of future directions, including the generalisation to massless higher spin fields and to de Sitter spacetime. Details of the derivation of the results shown here will be presented in a companion paper \cite{MStoappear}.

\paragraph*{Set-up}
We work in Euclidean signature and write the full metric, $\bar g$, as a sum of the background metric, $g$, and a perturbation around it, $h$. Our starting point will be the BRST invariant action for $h$ obtained from the EH action for $\bar g$ by rewriting $\bar g_{\mu\nu}=g_{\mu\nu}+h_{\mu\nu}$. The full action is given by
\begin{equation}
    S[b,c,h;g]=S_{\rm EH}[h;g]+S_{\rm gf}[h;g]+S_{\rm gh}[b,c,h;g],
\end{equation}
where $h$ is the fluctuating graviton field, $b$ and $c$ are the antighost and ghost fields, the semicolon denotes that $g$ is a background field, and
\begin{widetext}
    \begin{align}
        S_{\rm EH}[h;g]&\equiv S_{\rm EH}[\bar g]=-\frac{1}{2\kappa^2}\int\mathrm{d}^{d+1}x\sqrt{\bar g} \, \left(\bar R-2\Lambda\right),\\
        S_{\rm gf}[h;g]&\equiv\frac{1}{\kappa^2}\int\mathrm{d}^{d+1}x\sqrt g\left[\frac 1\alpha F_\mu(h) F^\mu(h)\right], \quad F^\mu(h)=\left(g^{\mu\rho}\nabla^\nu-\beta g^{\nu\rho}\nabla^\mu\right)h_{\nu\rho}\label{gauge fixing},\\
        S_{\rm gh}[b,c,h;g]&\equiv\frac{1}{\kappa^2}\int\mathrm{d}^{d+1}x\sqrt g\left[b_\mu\left(g^{\mu\rho}\nabla^\nu-\beta g^{\nu\rho}\nabla^\mu\right)\left(\bar\nabla_\nu c_\rho+\bar\nabla_\rho c_\nu\right)\right]=\\
        &=\frac{1}{\kappa^2}\int\mathrm{d}^{d+1}x\sqrt g\left[b_\mu\left(g^{\mu\nu}\nabla^\rho\nabla_\rho+\nabla^\nu\nabla^\mu-2\beta\nabla^\mu\nabla^\nu\right)c_\nu+\mathcal{O}(h)\right]\, , \label{ghost}
    \end{align}
\end{widetext}
where we use the conventions in \cite{Polchinski:1998rq} for the BRST transformations. Here $\kappa$ is the gravitational coupling constant related to Newton's constant by $\kappa^2=8\pi G_N$, $\Lambda$ is the cosmological constant, $\bar\nabla$ and $\bar R$ are the covariant derivative and the Ricci scalar induced by the metric $\bar g$, and $\nabla$ is the covariant derivative induced by the metric $g$. The gauge-fixing function $F^\mu(h)$ in \eqref{gauge fixing} is the most general covariant gauge-fixing term one can write down for a spin-$2$ field \cite{Capper:1979ej}. Because it contains two terms, there are also two independent gauge fixing parameters, $\alpha$ and $\beta$.\footnote{This is unlike in the case of a spin-$1$ field for which the most general covariant gauge-fixing term is $1/(2\xi)\int\mathrm{d}^{d+1}x\sqrt{g}\left(\nabla^\mu A_\mu\right)^2$ and depends on a single gauge fixing parameter $\xi$.} In the ghost action $S_{\rm gh}[b,c,h;g]$ in \eqref{ghost} we separated the ghost kinetic term from the (infinite number of) interaction terms between the ghosts and the graviton. The full path integral reads
\begin{equation}
    Z=\int \mathcal{D}b\mathcal{D}c\mathcal{D}h\,e^{-S[b,c,h;g]},
\end{equation}
and can be used to determine the correlators of the theory. The free ghost and graviton propagators are defined to match the $\mathcal{O}(\kappa^2)$ term in the perturbative expansion of the corresponding two-point functions,
\begin{align}
    \left\langle c_\mu(x)b_{\nu'}(y)\right\rangle&=G_{\rm {gh},\mu\nu'}(x,y)+\mathcal{O}(\kappa^3),\\
    \left\langle h_{\mu\nu}(x)h_{\alpha'\beta'}(y)\right\rangle&=G_{\mu\nu,\alpha'\beta'}(x,y)+\mathcal{O}(\kappa^3),
\end{align}
where here and below we use unprimed indices to indicate tensors at the point $x$ and primed indices for tensors at the point $y$. The propagators obey the equations
\begin{multline}
    \left(g^{\mu\nu}\nabla^\rho\nabla_\rho+\nabla^\nu\nabla^\mu-2\beta\nabla^\mu\nabla^\nu\right)G_{\rm {gh},\nu\rho'}(x,y)=\\
    =\frac{\kappa^2}{\sqrt g}\delta^\mu_{\rho'}\delta^{(d+1)}(x-y),\label{ghost eq}
\end{multline}
\begin{multline}
    \mathcal{D}^{\mu\nu\alpha\beta}G_{\alpha\beta,\rho'\sigma'}(x,y)=\\
    =-\frac{\kappa^2}{\sqrt g}\left(\frac12\delta^\mu_{\sigma'}\delta^\nu_{\rho'}+\frac12\delta^\mu_{\rho'}\delta^\nu_{\sigma'}\right)\delta^{(d+1)}(x-y),\label{graviton eq}
\end{multline}
where $\mathcal{D}^{\mu\nu\alpha\beta}$ is the second-order differential operator that depends on the background metric $g$, the cosmological constant $\Lambda$, and the gauge fixing parameters $\alpha$ and $\beta$, that appears in the action as
\begin{equation}
    \int\mathrm{d}^{d+1}x\sqrt g\left[-\frac1{2\kappa^2} h_{\mu\nu}\mathcal{D}^{\mu\nu\alpha\beta}h_{\alpha\beta}\right]\subset S[b,c,h;g].
\end{equation}
It is straightforward but tedious to work out $\mathcal{D}^{\mu\nu\alpha\beta}$. The explicit expression is too long and it will not be reported here (see, for example, \cite{Barth:1983hb} for useful formulae). In addition to equations \eqref{ghost eq} and \eqref{graviton eq} that the ghost and graviton propagators have to satisfy independently, BRST invariance implies an additional constraint equation \cite{Frob:2017gez,Glavan:2022nrd,Moga:2025gdy} due to the fact that the expectation value of any BRST-exact object must vanish. Specifically, we have that
\begin{equation}
    0=\left\langle\delta_{\rm {BRST}}\left(b_\mu(x)h_{\nu'\rho'}(y)\right)\right\rangle
\end{equation}
implies that
\begin{multline}
    \left(\delta_\mu^\beta\nabla^\alpha-\beta g^{\alpha\beta}\nabla_\mu\right)G_{\alpha\beta,\nu'\rho'}(x,y)=\\
    =\frac{\alpha}{2}\left(\nabla_{\nu'}G_{\rm {gh},\rho'\mu}(y,x)+\nabla_{\rho'}G_{\rm {gh},\nu'\mu}(y,x)\right).\label{brst eq}
\end{multline}

We will be looking for covariant expressions for the propagators and, thus, we make the following Ansätze\footnote{It is also possible to rewrite the basis of bitensors using only factors with one derivative of $\mu$ and replacing the factors with two derivatives of $\mu$ with the metric $g_{\mu\nu}$ and the parallel propagator $g_{\mu\alpha'}$ using $\nabla_\mu\nabla_\nu\mu=\left(g_{\mu\nu}-\nabla_\mu\mu\nabla_\nu\mu\right)/\tanh(\mu)$ and $\nabla_\mu\nabla_{\alpha'}\mu=-\left(g_{\mu\alpha'}+\nabla_\mu\mu\nabla_{\alpha'}\mu\right)/\sinh(\mu)$.}
\begin{align}
    &G_{\rm {gh},\mu\nu'}(x,y)=A(\mu) \nabla_{\nu' }\nabla_{\mu }\mu + B(\mu) \nabla_{\mu }\mu \nabla_{\nu' }\mu,\label{ghost ansatz}\\
    &G_{\mu\nu,\alpha'\beta'}(x,y)=\nonumber\\
    &=C_1(\mu) (\nabla_{\alpha'}\nabla_{\nu }\mu \nabla_{\beta'}\nabla_{\mu }\mu + \nabla_{\alpha'}\nabla_{\mu }\mu \nabla_{\beta'}\nabla_{\nu }\mu)\nonumber\\
    & + C_2(\mu) (\nabla_{\nu }\mu \nabla_{\alpha'}\nabla_{\mu }\mu \nabla_{\beta'}\mu + \nabla_{\mu }\mu \nabla_{\alpha'}\nabla_{\nu }\mu \nabla_{\beta'}\mu\nonumber\\
    &\quad\quad\quad + \nabla_{\nu }\mu \nabla_{\alpha'}\mu \nabla_{\beta'}\nabla_{\mu }\mu + \nabla_{\mu }\mu \nabla_{\alpha'}\mu \nabla_{\beta'}\nabla_{\nu }\mu)\nonumber\\
    & + C_3(\mu) \nabla_{\mu }\nabla_{\nu }\mu \nabla_{\alpha'}\nabla_{\beta'}\mu\label{graviton ansatz}\\
    & + C_4(\mu) (\nabla_{\mu }\mu \nabla_{\nu }\mu \nabla_{\alpha'}\nabla_{\beta'}\mu + \nabla_{\mu }\nabla_{\nu }\mu \nabla_{\alpha'}\mu \nabla_{\beta'}\mu)\nonumber\\
    & + C_5(\mu) \nabla_{\mu }\mu \nabla_{\nu }\mu \nabla_{\alpha'}\mu \nabla_{\beta'}\mu\nonumber,
\end{align}
where $\mu(x,y)$ is the geodesic distance between the points $x$ and $y$. The choice of this basis of bitensors for the propagators was made based on the experience with the photon propagator computed in \cite{Moga:2025gdy}. For details on how to manipulate bitensors we refer the reader to the original work by Allen and Jacobson \cite{Allen:1985wd}.

The discussion so far has been completely general and holds for any background metric $g_{\mu\nu}$. We will now specialize to (Euclidean) AdS$_{d+1}$. The system of equations governing $A,B,C_i, \, i=1,2,3,4,5,$ is intractable for general $(\alpha, \beta, d)$. The equations can be completely integrated for a number of special values of the triplet $(\alpha, \beta, d)$: (i) $(\alpha, 5/2, 3)$ (a dS version of this solution has been obtained earlier in \cite{Higuchi:2001uv}), (ii) $(36/5, 7/10, 3)$, (iii) $(\alpha, 2/3, 4)$. These solutions will be presented in \cite{MStoappear}.  

Significant simplifications happen when $\beta=1$. This value is special in a number of ways. First, for this value of $\beta$ and for a flat background, $g_{\mu\nu}=\delta_{\mu\nu}$, the gauge-fixing function does not completely fix the gauge, since $F^\alpha(h_{\mu \nu}{=}\partial_\mu \partial_\nu \lambda)=0$ identically for any scalar function $\lambda$. This can also be seen from the quadratic part of the gauge-fixed action. In flat spacetime and after commuting the derivatives the ghost kinetic term in \eqref{ghost} becomes
\begin{equation}
    \frac{1}{\kappa^2}\int\mathrm{d}^{d+1}x\left[b_\mu\left(\delta^{\mu\nu}\partial^\rho\partial_\rho+(1-2\beta)\partial^\mu\partial^\nu\right)c_\nu\right],
\end{equation}
which for $\beta=1$ is nothing but the kinetic term of the usual Maxwell action (up to a total derivative). Because of the gauge symmetry of the Maxwell action, this kinetic term has zero modes and is not invertible. The same is true for the graviton kinetic term (since the gauge-fixing terms did not fix the gauge completely). Hence, for $\beta=1$ the propagators cannot be defined in flat spacetime (without further gauge fixing this invariance). This is, however, not true in curved spacetimes as commuting the derivatives introduces curvature terms and the kinetic terms remain invertible even when $\beta=1$. In particular, for maximally symmetric spacetimes the curvature terms introduced are proportional to the metric and act as effective masses for the fields. Thus, interestingly, it appears that there are gauge choices for the propagators in curved spacetimes that are not possible in flat spacetime! When $\beta=1$ the equations are solvable for (i) arbitrary $\alpha$ given integer $d$ and (ii) $\alpha=4(d+2)/d$ with arbitrary $d$.\footnote{Interestingly, up to an overall factor of $4$ which can be absorbed in the normalization of the gauge-fixing function $F^\mu(h)$ in \eqref{gauge fixing}, this value of $\alpha$ is the same as the value of $\xi$ for the Fried-Yennie gauge of a photon (see below) in two dimensions higher \cite{Moga:2025gdy}.} The solutions are simplest in this last case, and this is the focus of this Letter (the other cases will be discussed in \cite{MStoappear}).

\paragraph*{Results}
We now present the results for this gauge choice.  We set the AdS radius equal to unity, so the cosmological constant is $\Lambda=-d(d-1)/2$. The Ricci tensor and Ricci scalar also become $R_{\mu\nu}=-d g_{\mu\nu}$ and $R=-d(d+1)$. To solve for the propagators we follow the methodology presented in \cite{Moga:2025gdy}. We plug our Ansätze \eqref{ghost ansatz} and \eqref{graviton ansatz} in the ghost and graviton propagator equations \eqref{ghost eq} and \eqref{graviton eq} as well as in the BRST constraint equation \eqref{brst eq}. Simplifying one obtains a set of coupled ordinary differential equations for the form factors $A,B,C_i, \, i=1,2,3,4,5$. Since the set of equations one obtains from the graviton propagator equation \eqref{graviton eq} for the form factors $C_i, \, i=1,2,3,4,5,$ is complicated to solve, the idea is to first use the ghost propagator equation \eqref{ghost eq} to solve for $A$ and $B$, and then consider the system of equations arising from the BRST constraint equation \eqref{brst eq} and the graviton propagator equation \eqref{graviton eq} together. This system of equations is overdetermined, but, at least in some cases, this makes it solvable as one can use substitution repeatedly to decouple the equations, express all of the form factors in terms of just one of them, and obtain a not-so-complicated ordinary differential equation for the final form factor.

Following this procedure, we find an extremely simple solution for the graviton propagator in this special covariant gauge:\footnote{The expressions for the ghost propagator reduce to elementary functions for given integer values of $d$ (similarly to what happens in the spin-$1$ case). When $d$ is even they are given by polynomials of hyperbolic functions, whereas when $d$ is odd there are also logarithms of hyperbolic functions.}
\begin{align}
    A(\mu)&=\kappa^2 \frac {\Gamma\left (\frac {d + 
       3} 2\right)\sinh(\mu)} {\pi^{\frac {d + 1} 2}\cosh^{d+1}(\mu)}\\
    &\times\Bigg(\frac{1}{d(d+2)} \, _2F_1\left(\frac{d+2}{2},\frac{d+3}{2};\frac{d+4}{2};\text{sech}^2(\mu )\right)\nonumber\\
    &-\frac{\text{sech}^2(\mu )}{2d(d+4)} \,_2F_1\left(\frac{d+3}{2},\frac{d+4}{2};\frac{d+6}{2};\text{sech}^2(\mu )\right)\Bigg),\nonumber\\
    B(\mu)&=\frac{A'(\mu)}{\cosh(2\mu)},\\
    C_1(\mu)
    &=\kappa^2 \frac {\Gamma\left (\frac {d + 
       3} 2\right)} {\pi^{\frac {d + 1} 2}\sinh^d(\mu)}\frac{\sinh(2\mu)}{d(d-1)},\\
    C_2(\mu)&=\kappa^2 \frac {\Gamma\left (\frac {d + 
       3} 2\right)} {\pi^{\frac {d + 1} 2}\sinh^d(\mu)}\frac{-(d+2)}{d^2},\\
    C_3(\mu)&=\kappa^2 \frac {\Gamma\left (\frac {d + 
       3} 2\right)} {\pi^{\frac {d + 1} 2}\sinh^d(\mu)}\frac{-4\tanh(\mu)}{d^2(d-1)},\\
    C_4(\mu)&= C_5(\mu)=  0.
\end{align}
The solution obtained satisfies the ghost and graviton propagator equations \eqref{ghost eq} and \eqref{graviton eq} and the BRST constraint equation \eqref{brst eq}. The integration constants were fixed by demanding that at large separation, $\mu\to\infty$, the propagators vanish (at the rate of a normalizable mode) and that at short distance, $\mu\to0$, the solutions reproduce the delta-function constraints on the right-hand side of the ghost and graviton propagator equations \eqref{ghost eq} and \eqref{graviton eq}. 

If desired, it is straightforward to rewrite the propagators in other commonly used AdS-invariant variables such as $u=\cosh \mu-1$ or the chordal distance $\xi=1/(1+u)$, in which the form factors are rational functions. The solution also reproduces the gauge independent parts of the propagator obtained in \cite{DHoker:1999bve}. As discussed there, these parts are given by hypergeometric functions and are determined in terms of the propagator of a massless scalar, which itself is given by a hypergeometric function. Changing our basis to theirs yields differential relations between our (simpler) form factors and theirs and integrating these relations reproduces their results.

It is remarkable how simple the expression for the graviton propagator in arbitrary dimensions is in this gauge compared to any other previously reported solutions in the literature. It is also worth noting that, as alluded to in the introduction, the simplicity of the propagator in this gauge is the result of some cancellation between its gauge independent and gauge dependent parts, none of which is simple on its own. This dramatic simplification of the propagator in this gauge is reminiscent of the simplification of the photon propagator in the Fried-Yennie gauge that was recently discussed in both dS \cite{Glavan:2022nrd,Glavan:2022dwb} and AdS \cite{Ciccone:2024guw,Moga:2025gdy}. In fact, as discussed in detail in \cite{Moga:2025gdy}, in flat spacetime the Fried-Yennie gauge for the position space propagator is analogous to the momentum space Landau propagator. The latter obeys the relation $p^\mu G_{\mu\nu'}(p)=0$ because of which it exhibits an improved ultraviolet behavior. In the same manner, the position space propagator in the Fried-Yennie gauge obeys the relation $\nabla^\mu\mu \, G_{\mu\nu'}(\mu)=0$ and it exhibits an improved IR behaviour. The graviton propagator found here obeys the natural generalization of this relation,
\begin{equation}
    \nabla^\mu\mu\nabla^\nu\mu \, G_{\mu\nu,\alpha'\beta'}(\mu)=0.
\end{equation}
The choice of our basis of bitensors in our Ansätze was made precisely because it makes manifest this relation in this gauge.\footnote{For an arbitrary gauge with generic $\alpha$ and $\beta$ we have
\begin{equation}
    \nabla^\mu\mu\nabla^\nu\mu \, G_{\mu\nu,\alpha'\beta'}(\mu)=C_4(\mu)\nabla_{\alpha'}\nabla_{\beta'}\mu+C_5(\mu)\nabla_{\alpha'}\mu\nabla_{\beta'}\mu.
\end{equation}
} It is worth noting that propagators in AdS take simpler forms in gauges with improved IR behaviour, while at the same time AdS can be considered as an IR regulator of theories in flat spacetime \cite{Callan:1989em}. Note, however, that $\beta=1$ is precisely a gauge choice that does not have a flat spacetime counterpart. On a similar note, if one tried to look for solutions for the flat spacetime propagator that have $C_4=C_5=0$, one would be forced into choosing complex values for the gauge fixing parameters $\alpha$ and $\beta$.

Based on these results, we suggest that there exists a covariant gauge for spin-$s$ massless fields for which the propagator obeys
\begin{equation}
    \nabla^{\mu_1}\mu\ldots\nabla^{\mu_s}\mu \, G_{\mu_1\ldots\mu_s,\alpha'_1\ldots\alpha'_s}(\mu)=0,
\end{equation}
and in which it takes a generally simple form. A natural guess for the generalisation of the formulas for the special values of the gauge fixing parameters $\xi=d/(d-2)$ (spin-$1$) and $\alpha/4=(d+2)/d$ (spin-$2$) is that the gauge-parameter (for suitably normalized gauge-fixing term and suitable choice of any other gauge fixing parameters) is $\xi_s=(d+2(s-1))/(d-2+2(s-1))$. It would be interesting to investigate further the spin-$s$ case.

\paragraph*{Conclusions}
In this Letter we presented a remarkably simple expression for the graviton propagator in a special covariant gauge in AdS. The simplicity of the propagator in this gauge is linked to an improved IR behaviour. We have suggested that a similar gauge may also exist for higher spin gauge fields. The expression for the propagator should make tractable the evaluation of Witten-Feynman diagrams in AdS, which would otherwise be out of reach, and since the propagators are valid in arbitrary dimensions using dimensional regularization to regulate the corresponding integrals is a viable choice. Another generalisation is to the case of de Sitter spacetime. It is simple to see (and we have checked this by an independent computation) that there is a corresponding dS solution of the equations defining the propagators obtained by taking $\mu \to i \mu$ in our AdS solution. The physical relevance of this solution requires further analysis. We hope to return to these and related issues in future work.

\paragraph*{Acknowledgements}
R.M. is supported by an STFC studentship. K.S. is supported in part by the STFC consolidated grant ST/X000583/1 ``New Frontiers in Particle Physics, Cosmology and Gravity." 

\bibliographystyle{apsrev}
\bibliography{refs}

@book{Polchinski:1998rq,
    author = "Polchinski, J.",
    title = "{String theory. Vol. 1: An introduction to the bosonic string}",
    doi = "10.1017/CBO9780511816079",
    isbn = "978-0-511-25227-3, 978-0-521-67227-6, 978-0-521-63303-1",
    publisher = "Cambridge University Press",
    series = "Cambridge Monographs on Mathematical Physics",
    month = "12",
    year = "2007"
}

@article{Tsamis:1992zt,
    author = "Tsamis, N. C. and Woodard, R. P.",
    title = "{Mode analysis and ward identities for perturbative quantum gravity in de Sitter space}",
    reportNumber = "UFIFT-HEP-92-20, CRETE-92-13",
    doi = "10.1016/0370-2693(92)91174-8",
    journal = "Phys. Lett. B",
    volume = "292",
    pages = "269--276",
    year = "1992"
}

@article{Moga:2025gdy,
    author = "Moga, Radu N. and Skenderis, Kostas",
    title = "{Bulk-to-bulk photon propagator in AdS}",
    eprint = "2510.23770",
    archivePrefix = "arXiv",
    primaryClass = "hep-th",
    doi = "10.1007/JHEP05(2026)205",
    journal = "JHEP",
    volume = "05",
    pages = "205",
    year = "2026"
}

@article{Higuchi:2011vw,
    author = "Higuchi, Atsushi and Marolf, Donald and Morrison, Ian A.",
    title = "{de Sitter invariance of the dS graviton vacuum}",
    eprint = "1107.2712",
    archivePrefix = "arXiv",
    primaryClass = "hep-th",
    doi = "10.1088/0264-9381/28/24/245012",
    journal = "Class. Quant. Grav.",
    volume = "28",
    pages = "245012",
    year = "2011"
}

@article{Morrison:2013rqa,
    author = "Morrison, Ian A.",
    title = "{On cosmic hair and ''de Sitter breaking'' in linearized quantum gravity}",
    eprint = "1302.1860",
    archivePrefix = "arXiv",
    primaryClass = "gr-qc",
    month = "2",
    year = "2013"
}

@article{Miao:2013isa,
    author = "Miao, S. P. and Mora, P. J. and Tsamis, N. C. and Woodard, R. P.",
    title = "{Perils of analytic continuation}",
    eprint = "1306.5410",
    archivePrefix = "arXiv",
    primaryClass = "gr-qc",
    reportNumber = "UFIFT-QG-13-03",
    doi = "10.1103/PhysRevD.89.104004",
    journal = "Phys. Rev. D",
    volume = "89",
    number = "10",
    pages = "104004",
    year = "2014"
}

@article{Glavan:2022nrd,
    author = "Glavan, Dra{\v{z}}en and Prokopec, Tomislav",
    title = "{Even the photon propagator must break de Sitter symmetry}",
    eprint = "2212.13997",
    archivePrefix = "arXiv",
    primaryClass = "hep-th",
    doi = "10.1016/j.physletb.2023.137928",
    journal = "Phys. Lett. B",
    volume = "841",
    pages = "137928",
    year = "2023"
}

@article{Glavan:2022dwb,
    author = "Glavan, Dra{\v{z}}en and Prokopec, Tomislav",
    title = "{Photon propagator in de Sitter space in the general covariant gauge}",
    eprint = "2212.13982",
    archivePrefix = "arXiv",
    primaryClass = "gr-qc",
    doi = "10.1007/JHEP05(2023)126",
    journal = "JHEP",
    volume = "05",
    pages = "126",
    year = "2023"
}

@article{Frob:2017gez,
    author = {Fr{\"o}b, Markus B. and Taslimi Tehrani, Mojtaba},
    title = "{Green{\textquoteright}s functions and Hadamard parametrices for vector and tensor fields in general linear covariant gauges}",
    eprint = "1708.00444",
    archivePrefix = "arXiv",
    primaryClass = "gr-qc",
    doi = "10.1103/PhysRevD.97.025022",
    journal = "Phys. Rev. D",
    volume = "97",
    number = "2",
    pages = "025022",
    year = "2018"
}

@article{Ciccone:2024guw,
    author = "Ciccone, Riccardo and De Cesare, Fabiana and Di Pietro, Lorenzo and Serone, Marco",
    title = "{Exploring confinement in Anti-de Sitter space}",
    eprint = "2407.06268",
    archivePrefix = "arXiv",
    primaryClass = "hep-th",
    doi = "10.1007/JHEP12(2024)218",
    journal = "JHEP",
    volume = "12",
    pages = "218",
    year = "2024",
    note = "[Erratum: JHEP 06, 037 (2025)]"
}

@article{Callan:1989em,
    author = "Callan, Jr., Curtis G. and Wilczek, Frank",
    title = "{INFRARED BEHAVIOR AT NEGATIVE CURVATURE}",
    reportNumber = "IASSNS-HEP-90-4, PUPT-1168",
    doi = "10.1016/0550-3213(90)90451-I",
    journal = "Nucl. Phys. B",
    volume = "340",
    pages = "366--386",
    year = "1990"
}

@article{Feynman:1963ax,
    author = "Feynman, R. P.",
    editor = "Hsu, Jong-Ping and Fine, D.",
    title = "{Quantum theory of gravitation}",
    journal = "Acta Phys. Polon.",
    volume = "24",
    pages = "697--722",
    year = "1963"
}

@article{DeWitt:1967yk,
    author = "DeWitt, Bryce S.",
    editor = "Fang, Li-Zhi and Ruffini, R.",
    title = "{Quantum Theory of Gravity. 1. The Canonical Theory}",
    doi = "10.1103/PhysRev.160.1113",
    journal = "Phys. Rev.",
    volume = "160",
    pages = "1113--1148",
    year = "1967"
}

@article{DeWitt:1967ub,
    author = "DeWitt, Bryce S.",
    editor = "Hsu, Jong-Ping and Fine, D.",
    title = "{Quantum Theory of Gravity. 2. The Manifestly Covariant Theory}",
    doi = "10.1103/PhysRev.162.1195",
    journal = "Phys. Rev.",
    volume = "162",
    pages = "1195--1239",
    year = "1967"
}

@article{DeWitt:1967uc,
    author = "DeWitt, Bryce S.",
    editor = "Hsu, Jong-Ping and Fine, D.",
    title = "{Quantum Theory of Gravity. 3. Applications of the Covariant Theory}",
    doi = "10.1103/PhysRev.162.1239",
    journal = "Phys. Rev.",
    volume = "162",
    pages = "1239--1256",
    year = "1967"
}

@article{tHooft:1974toh,
    author = "'t Hooft, Gerard and Veltman, M. J. G.",
    title = "{One-loop divergencies in the theory of gravitation}",
    doi = "10.1142/9789814539395_0001",
    journal = "Ann. Inst. H. Poincare Phys. Theor. A",
    volume = "20",
    number = "1",
    pages = "69--94",
    year = "1974"
}

@article{Maldacena:1997re,
    author = "Maldacena, Juan Martin",
    title = "{The Large $N$ limit of superconformal field theories and supergravity}",
    eprint = "hep-th/9711200",
    archivePrefix = "arXiv",
    reportNumber = "HUTP-97-A097, HUTP-98-A097",
    doi = "10.4310/ATMP.1998.v2.n2.a1",
    journal = "Adv. Theor. Math. Phys.",
    volume = "2",
    pages = "231--252",
    year = "1998"
}

@article{Gubser:1998bc,
    author = "Gubser, S. S. and Klebanov, Igor R. and Polyakov, Alexander M.",
    title = "{Gauge theory correlators from noncritical string theory}",
    eprint = "hep-th/9802109",
    archivePrefix = "arXiv",
    reportNumber = "PUPT-1767",
    doi = "10.1016/S0370-2693(98)00377-3",
    journal = "Phys. Lett. B",
    volume = "428",
    pages = "105--114",
    year = "1998"
}

@article{Witten:1998qj,
    author = "Witten, Edward",
    title = "{Anti de Sitter space and holography}",
    eprint = "hep-th/9802150",
    archivePrefix = "arXiv",
    reportNumber = "IASSNS-HEP-98-15",
    doi = "10.4310/ATMP.1998.v2.n2.a2",
    journal = "Adv. Theor. Math. Phys.",
    volume = "2",
    pages = "253--291",
    year = "1998"
}

@article{Allen:1986ta,
    author = "Allen, Bruce",
    title = "{The Graviton Propagator in De Sitter Space}",
    reportNumber = "TUTP-86-9",
    doi = "10.1103/PhysRevD.34.3670",
    journal = "Phys. Rev. D",
    volume = "34",
    pages = "3670",
    year = "1986"
}

@article{Allen:1986tt,
    author = "Allen, Bruce and Turyn, Michael",
    title = "{An Evaluation of the Graviton Propagator in De Sitter Space}",
    reportNumber = "TUTP-86-20",
    doi = "10.1016/0550-3213(87)90672-9",
    journal = "Nucl. Phys. B",
    volume = "292",
    pages = "813",
    year = "1987"
}

@article{Antoniadis:1986sb,
    author = "Antoniadis, Ignatios and Mottola, E.",
    title = "{Graviton Fluctuations in De Sitter Space}",
    reportNumber = "CERN-TH-4605/86",
    doi = "10.1063/1.529381",
    journal = "J. Math. Phys.",
    volume = "32",
    pages = "1037--1044",
    year = "1991"
}

@article{Higuchi:2001uv,
    author = "Higuchi, Atsushi and Kouris, Spyros S.",
    title = "{The Covariant graviton propagator in de Sitter space-time}",
    eprint = "gr-qc/0107036",
    archivePrefix = "arXiv",
    doi = "10.1088/0264-9381/18/20/311",
    journal = "Class. Quant. Grav.",
    volume = "18",
    pages = "4317--4328",
    year = "2001"
}

@article{Miao:2011fc,
    author = "Miao, S. P. and Tsamis, N. C. and Woodard, R. P.",
    title = "{The Graviton Propagator in de Donder Gauge on de Sitter Background}",
    eprint = "1106.0925",
    archivePrefix = "arXiv",
    primaryClass = "gr-qc",
    reportNumber = "ITP-UU-11-18, SPIN-11-13, CCTP-2011-15, UFIFT-QG-11-03",
    doi = "10.1063/1.3664760",
    journal = "J. Math. Phys.",
    volume = "52",
    pages = "122301",
    year = "2011"
}

@article{Kahya:2011sy,
    author = "Kahya, E. O. and Miao, S. P. and Woodard, R. P.",
    title = "{The Coincidence Limit of the Graviton Propagator in de Donder Gauge on de Sitter Background}",
    eprint = "1112.4420",
    archivePrefix = "arXiv",
    primaryClass = "gr-qc",
    reportNumber = "UFIFT-QG-11-05",
    doi = "10.1063/1.3681886",
    journal = "J. Math. Phys.",
    volume = "53",
    pages = "022304",
    year = "2012"
}

@article{Mora:2012zi,
    author = "Mora, P. J. and Tsamis, N. C. and Woodard, R. P.",
    title = "{Graviton Propagator in a General Invariant Gauge on de Sitter}",
    eprint = "1205.4468",
    archivePrefix = "arXiv",
    primaryClass = "gr-qc",
    reportNumber = "CCTP-2012-09, UFIFT-QC-12-05",
    doi = "10.1063/1.4764882",
    journal = "J. Math. Phys.",
    volume = "53",
    pages = "122502",
    year = "2012"
}

@article{Frob:2016hkx,
    author = {Fr{\"o}b, Markus B. and Higuchi, Atsushi and Lima, William C. C.},
    title = "{Mode-sum construction of the covariant graviton two-point function in the Poincar{\'e} patch of de Sitter space}",
    eprint = "1603.07338",
    archivePrefix = "arXiv",
    primaryClass = "gr-qc",
    doi = "10.1103/PhysRevD.93.124006",
    journal = "Phys. Rev. D",
    volume = "93",
    number = "12",
    pages = "124006",
    year = "2016"
}

@article{Tsamis:1992xa,
    author = "Tsamis, N. C. and Woodard, R. P.",
    title = "{The Structure of perturbative quantum gravity on a De Sitter background}",
    reportNumber = "UFIFT-HEP-92-14, CRETE-92-11",
    doi = "10.1007/BF02102015",
    journal = "Commun. Math. Phys.",
    volume = "162",
    pages = "217--248",
    year = "1994"
}

@article{Higuchi:2002sc,
    author = "Higuchi, Atsushi and Weeks, Richard H.",
    title = "{The Physical graviton two point function in de Sitter space-time with S3 spatial sections}",
    eprint = "gr-qc/0212031",
    archivePrefix = "arXiv",
    doi = "10.1088/0264-9381/20/14/303",
    journal = "Class. Quant. Grav.",
    volume = "20",
    pages = "3005--3022",
    year = "2003"
}

@article{Glavan:2019msf,
    author = "Glavan, D. and Miao, S. P. and Prokopec, T. and Woodard, R. P.",
    title = "{Graviton Propagator in a 2-Parameter Family of de Sitter Breaking Gauges}",
    eprint = "1908.06064",
    archivePrefix = "arXiv",
    primaryClass = "gr-qc",
    reportNumber = "UFIFT-QG-19-03, CP3-19-39",
    doi = "10.1007/JHEP10(2019)096",
    journal = "JHEP",
    volume = "10",
    pages = "096",
    year = "2019"
}

@article{Glavan:2025azq,
    author = "Glavan, Dra{\v{z}}en",
    title = "{Graviton propagator in de Sitter space in a simple one-parameter gauge}",
    eprint = "2511.13660",
    archivePrefix = "arXiv",
    primaryClass = "gr-qc",
    month = "11",
    year = "2025"
}

@article{Liu:1998ty,
    author = "Liu, Hong and Tseytlin, Arkady A.",
    title = "{On four point functions in the CFT / AdS correspondence}",
    eprint = "hep-th/9807097",
    archivePrefix = "arXiv",
    reportNumber = "IMPERIAL-TP-97-98-060",
    doi = "10.1103/PhysRevD.59.086002",
    journal = "Phys. Rev. D",
    volume = "59",
    pages = "086002",
    year = "1999"
}

@article{Raju:2011mp,
    author = "Raju, Suvrat",
    title = "{Recursion Relations for AdS/CFT Correlators}",
    eprint = "1102.4724",
    archivePrefix = "arXiv",
    primaryClass = "hep-th",
    reportNumber = "HRI-ST-1103",
    doi = "10.1103/PhysRevD.83.126002",
    journal = "Phys. Rev. D",
    volume = "83",
    pages = "126002",
    year = "2011"
}

@article{DHoker:1999bve,
    author = "D'Hoker, Eric and Freedman, Daniel Z. and Mathur, Samir D. and Matusis, Alec and Rastelli, Leonardo",
    title = "{Graviton and gauge boson propagators in AdS(d+1)}",
    eprint = "hep-th/9902042",
    archivePrefix = "arXiv",
    reportNumber = "MIT-CTP-2824, UCLA-99-TEP-1",
    doi = "10.1016/S0550-3213(99)00524-6",
    journal = "Nucl. Phys. B",
    volume = "562",
    pages = "330--352",
    year = "1999"
}

@article{Kolar:2023mkw,
    author = "Kol{\'a}{\v{r}}, Ivan and M{\'a}lek, Tom{\'a}{\v{s}}",
    title = "{Propagators in AdS for higher-derivative and nonlocal gravity: Heat kernel approach}",
    eprint = "2307.13056",
    archivePrefix = "arXiv",
    primaryClass = "gr-qc",
    doi = "10.1140/epjc/s10052-025-13769-y",
    journal = "Eur. Phys. J. C",
    volume = "85",
    number = "2",
    pages = "171",
    year = "2025"
}

@article{Allen:1985wd,
    author = "Allen, Bruce and Jacobson, Theodore",
    title = "{Vector Two Point Functions in Maximally Symmetric Spaces}",
    reportNumber = "UCSB-TH-4-1985",
    doi = "10.1007/BF01211169",
    journal = "Commun. Math. Phys.",
    volume = "103",
    pages = "669",
    year = "1986"
}

@article{Banados:2022nhj,
    author = "Ba{\~n}ados, M{\'a}ximo and Bianchi, Ernesto and Mu{\~n}oz, Iv{\'a}n and Skenderis, Kostas",
    title = "{Bulk renormalization and the AdS/CFT correspondence}",
    eprint = "2208.11539",
    archivePrefix = "arXiv",
    primaryClass = "hep-th",
    doi = "10.1103/PhysRevD.107.L021901",
    journal = "Phys. Rev. D",
    volume = "107",
    number = "2",
    pages = "L021901",
    year = "2023"
}

@article{MStoappear,
    author = "Moga, Radu N. and Skenderis, Kostas",
    journal = "to appear",
    year = "2026"
}

@article{Capper:1979ej,
    author = "Capper, D. M.",
    title = "{A GENERAL GAUGE GRAVITON LOOP CALCULATION}",
    reportNumber = "QMC-79-11",
    doi = "10.1088/0305-4470/13/1/022",
    journal = "J. Phys. A",
    volume = "13",
    pages = "199",
    year = "1980"
}

@article{Barth:1983hb,
    author = "Barth, N. H. and Christensen, S. M.",
    title = "{Quantizing Fourth Order Gravity Theories. 1. The Functional Integral}",
    reportNumber = "UNC-IFP-199",
    doi = "10.1103/PhysRevD.28.1876",
    journal = "Phys. Rev. D",
    volume = "28",
    pages = "1876",
    year = "1983"
}

\end{document}